\documentstyle[multicol,aps]{revtex}
\begin{document}

\title{Movement and fluctuations of the vacuum}
\author{Marc-Thierry Jaekel$^a$ and Serge Reynaud$^b$}
\address{$(a)$ Laboratoire de Physique Th\'{e}orique
de l'Ecole Normale Sup\'erieure, \\
Centre National de la Recherche Scientifique,
Universit\'e Paris-Sud, \\
24 rue Lhomond, F75231 Paris Cedex 05 France\\
$(b)$ Laboratoire Kastler Brossel,
Universit\'e Pierre et Marie Curie, \\
Ecole Normale Sup\'erieure,
Centre National de la Recherche Scientifique, \\
4 place Jussieu, F75252 Paris Cedex 05 France}
\date{LPTENS 97/22}
\maketitle

\begin{abstract}
Quantum fields possess zero-point or vacuum fluctuations
which induce mechanical effects, namely
generalised Casimir forces, on any scatterer.

Symmetries of vacuum therefore raise fundamental questions
when confronted with the principle of relativity
of motion in vacuum.
The specific case of uniformly accelerated motion is particularly
interesting, in connection with the much debated question
of the appearance of vacuum in accelerated frames.
The choice of Rindler representation, commonly used in General
Relativity, transforms vacuum fluctuations into thermal fluctuations,
raising difficulties of interpretation.
In contrast, the conformal representation of uniformly accelerated frames
fits the symmetry properties of field
propagation and quantum vacuum and thus leads to extend
the principle of relativity of motion to uniform accelerations.

Mirrors moving in vacuum with a non uniform acceleration are known
to radiate. The associated radiation reaction force is directly
connected to fluctuating forces felt by motionless mirrors
through fluctuation-dissipation relations.
Scatterers in vacuum undergo a quantum Brownian motion
which describes irreducible quantum fluctuations.
Vacuum fluctuations impose ultimate limitations on
measurements of position in space-time, and thus challenge
the very concept of space-time localisation
within a quantum framework.

For test masses greater than Planck mass, the ultimate limit in localisation
is determined by gravitational vacuum fluctuations.
Not only positions in space-time, but also geodesic distances,
behave as quantum variables, reflecting the necessary quantum
nature of an underlying geometry.

\end{abstract}

\section{Introduction}

The effects of vacuum fluctuations in microscopic physics
have played an essential role in the development of Quantum ElectroDynamics
and Quantum Field Theory.
Vacuum fluctuations also have an impact on macroscopic physics
as they give rise to mechanical forces.
{}Fluctuating forces and motional forces, which generalise
the static Casimir forces, have recently been thoroughly examined.
It has been shown that vacuum fluctuations affect the notion of movement
in vacuum and are directly related to the principle of relativity of motion.
They also impose fundamental limitations on localisation and,
hence, on the notion of space-time.

The aim of the present paper is to discuss questions connected to
these problems. Since a complete and consistent theory
of inertial and gravitational phenomena is still lacking,
these questions have not yet been satisfactorily answered.
We however intend to show that the fluctuations of quantum
vacuum have a profound impact on fundamental concepts of relativity and, in
particular, that they shed light on the interpretation of space-time
localisation in a quantum formalism. We also hope to convince readers
that the various physical effects discussed here will have to be considered
in any consistent theoretical frame to be developed in the future. In order
to put our purpose in perspective, we first briefly recall the historical
development of the concept of motion in vacuum.

The notions of void and movement have been associated since the foundation
of the atomistic school of philosophy by Leucippus and Democritus. More than
two thousand years ago, atoms and void were introduced as complementary and
necessary premises to the comprehension of movement. As outlined by
Russell \cite{Russell}, this was a logical requirement, more than a physical
one, which was intended to answer objections formulated by Zeno against the
very possibility of movement. The concept of void played a crucial role
since it was allowing motion to take place freely. In this respect, the
absence of friction in void was considered from the very beginning as a
preliminary requirement for any physics of movement. Although
a quite different conception of space, the Aritostelian one, prevailed for a
long time,
the ideas promoted by the ancient atomistic school were revived by the rise
of modern physics.

Galileo introduced the idea that motions with uniform velocity have the
same status as rest and that only changes of velocity have to be explained.
With this principle of relativity of motion, the absence of resistance to
uniform motion was recognized as a condition for the understanding of
motion. Shortly after Galileo's time, vacuum became a matter for physical
investigation with the experiments of Torricelli, Pascal and von Guericke.
Going further along the same directions, Newton was able to lay down the
fundamental laws of mechanics and gravitation. He emphasized that the
observed motions of celestial bodies could only be consistent with
negligible friction, so that they must be thought to occur in an empty space.
{}From the exponential decreasing of air density with altitude,
Newton estimated that interplanetary space was indeed practically free of
matter. Vacuum and the empty space of mechanics were thus identified as the
celebrated Newtonian absolute space \cite{Koyre}.
{}From a pragmatic point of view, vacuum could have been qualified at that time
as
what remains when matter has been removed from a given region in space.

Although the Newtonian conception raised severe difficulties which
were immediately noticed, it survived until the advent of Relativity Theory.
Meanwhile, empty space was promoted to the status of a
universal arena, not only for motion of matter, but also for propagation of
fields such as electromagnetic fields. As is well known, the key point of
the Einsteinian objection against Newtonian physics is that the properties
attributed to empty space by classical mechanics are not compatible with the
symmetry properties of Maxwell equations governing electromagnetism
\cite{Einstein}.
Building a definition of space-time events upon light propagation,
Einstein was able to restore the compatibility between mechanics
and electromagnetism \cite{EinsteinRelat}.

The discovery of field fluctuations led to a further enrichment of the
notion of vacuum. At any non zero temperature in particular, space is filled
with blackbody radiation. It was demonstrated by Einstein that scattering of
these thermal field fluctuations damp the motion of any scatterer to rest
while giving rise to Brownian fluctuations of position \cite
{EinsteinBrownian}. The close connection between fluctuations and
dissipation was thus noticed for the first time. It follows that a more
precise definition of vacuum has to be given as what remains when matter and
field have been removed from a given region in space.

After having introduced the concept of quanta in physics to build a theory
of blackbody radiation \cite{Planck00},
Planck modified his law to account for zero-point
fluctuations, that is field fluctuations persisting at zero temperature and
corresponding to an energy of half a quantum per mode \cite{Planck11}. The
status of these fluctuations, quite controversial in a classical formalism
\cite{EinsteinStern,Nernst16}, was more firmly established by
the quantum theory of electromagnetic field \cite{Dirac}.
Heisenberg inequalities entail a profound modification of the notion of
movement since they forbid absolute rest. Position and velocity of a body
cannot be simultaneously and precisely fixed and thus possess unsurpassable
residual fluctuations. This property subsists in Quantum Field Theory where
field modes, each of which is equivalent to an harmonic oscillator, also
possess irreducible quantum fluctuations. The fundamental state of the
field, defined by the absence of any field excitation, still contains
zero-point field fluctuations, also called vacuum fluctuations. The vacuum
state may be identified with quantum empty space \cite{Sciama}. The principle
of
relativity of motion in quantum vacuum holds for uniform velocity, as a
consequence of Lorentz invariance of vacuum fluctuations \cite{Boyer73}.

The previous paragraphs sketch a succession of developments which
may be compared with the logical line of thought raised by
ancient atomism. Movement has to be defined with respect to something. When
void is considered as reference for movement, it has thus to be distinguished
from nothingness. But if something fills vacuum, it is not clear how
it can manage to allow motion without friction. As we have seen, modern
physics finally came to a logical answer to this paradox. Vacuum is
indeed filled with field fluctuations but these fluctuations do not oppose
uniform motion as a consequence of a fundamental symmetry of physics.
Lorentz invariance is not only a symmetry of the laws of physics, but also
of the vacuum state, that is the quantum version of empty space. The
Galilean principle of relativity of motion, that is the preservation of
uniform motion in vacuum, has finally been made compatible with a physical
description of vacuum. However, inertial and gravitational phenomena still
raise
essential questions.

The Newtonian laws of inertia deal with accelerated motion and not only
with uniform velocity. In order to discuss these laws in the light of the
previous arguments, one has to ask two related questions. What is the
appearance of vacuum fluctuations for an accelerated body? Does vacuum
oppose to accelerated motion? These questions remain nowadays a matter of
controversy and will be examined in the present paper.
Although there is only one relativistic definition of uniformly
accelerated motion, there is a variety of possible representations of
uniformly accelerated frames. With the common choice of Rindler
representation \cite{Rindler}, vacuum is found to be transformed into a
thermal bath with a temperature proportional to acceleration and to
Planck constant \cite{Davies75,Unruh76,BirrellDavies,Fulling}. This
correspondance between acceleration and temperature has
apparently been easily accepted because of its association with the
most spectacular effect predicted by quantum field theory in curved
spacetime, namely thermal particle creation due to curvature
\cite{ZeldovichStarobinsky,Hawking75}.
It is nevertheless clear that accelerated frames and curved spacetime are
completely different physical problems, from the point of view of general
relativity.

Decisive progress were brought by explicit computations of scattering of
quantum fields by perfect mirrors moving in vacuum \cite{Moore70,deWitt75}.
Mirrors
moving in vacuum with a non uniform acceleration were predicted to radiate
\cite{FullingDavies}. In the historical context recalled previously, this
means that motion in quantum vacuum gives rise to a friction force. It has
however to be emphasized that this force vanishes in the specific cases of
uniform velocity or uniform acceleration. This result was understood as
challenging the alleged equivalence between accelerated vacuum and thermal
bath. Although it remains logically compatible with it,
it points to the difficulty of finding an uncontested
physical significance to this equivalence.
This is not so surprising if one realises that Rindler transformations
do not preserve Maxwell laws. After such a transformation, light rays appear
to be curved while frequencies undergo a redshift during propagation
\cite{EinsteinAccel}. Moreover, uniform velocity and rest do not play the
same role after these transformations. In other words, the Rindler
representation
of accelerated frames significantly departs from the point of view of
invariance
and symmetry properties which made the success of Lorentz transformations
\cite{EinsteinRelat}. This point of view may however be applied also for
uniformly accelerated frames provided the latter are represented through
conformal coordinate tranformations \cite{Bateman09,Cunningham09} which fit
the symmetry properties of electromagnetic theory and of quantum vacuum.

The existence of vacuum field fluctuations has been recognised for a long time
and their effects thoroughly studied \cite{ItzyksonZuber}. The
possibility of manipulating vacuum fluctuations and, even, to squeeze their
effect on quantum noise in optical measurements has been demonstrated
\cite{LoudonKnight,KimbleWalls,GiacobinoFabre,ReynaudHeidmann}.
As far as macroscopic effects of vacuum fluctuations are concerned,
the investigation of static Casimir forces
\cite{Casimir48,PlunienMuller,MostepanenkoTrunov}
has led to an experimental demonstration of vacuum radiation pressure
\cite{DeriaginAbrikosova,Sparnaay58,Tabor68,Sabisky73,Lamoreaux97}.
The closely related effect of attraction of an atom to a conducting plate
has also been demonstrated \cite{SukenikBoshier}. However, vacuum
fluctuations remain a matter of debate mainly because their energy is
infinite. More strikingly, their energy per unit volume is infinite. This
problem may be considered as a particular occurence of well known formal
difficulties of Quantum Field Theory. It raises a
specific question concerning gravitational theory since energy is the source
of gravity in General Relativity \cite{EinsteinGR}. As most
situations within Quantum Field Theory only involve differences of energy,
vacuum energy may be considered to be zero by definition.
The energy of field states is then determined relatively to
vacuum.
In fact, this procedure does not answer the question. The
Casimir energy, which is a difference between energies of two vacuum
configurations, has to contribute to gravity. It is also known that the
definition of vacuum depends on the choice of coordinates and
does not meet the covariance requirement of General
Relativity \cite{BirrellDavies}.

It has sometimes been argued that these problems can be solved, at least from a
pragmatical point of view, if the microscopic and macroscopic domains are
clearly separated \cite{Rosenfeld63}. Taken seriously, this argument means
that fundamental laws of physics hold at the elementary level of microscopic
physics where we have to renounce to general principles of mechanics, while
the mechanical description of nature has a validity restricted to the
macroscopic
domain. This argument is known to be conceptually inconsistent and to
endanger the consistency of quantum formalism because it mixes too crudely
quantum and classical descriptions \cite{deWitt62}. It has now lost its
pragmatical
pertinence, due to the progress towards highly sensitive measurements of
macroscopic objects \cite{BraginskyKhalili,BockoOnofrio}. Since the quantum
level of sensitivity is approached for such measurements, it appears more and
more uneasy to delineate any {\it a priori} frontier between microscopic and
macroscopic domains. Moreover, the fluctuations of the stress
tensor corresponding to quantum matter and quantum field present in
space-time must induce fluctuations of the space-time metric itself \cite
{Blohincev60}. These arguments contribute to render a classical conception
of space-time untenable in the presence of quantum fields and plead
for a definition of localisation accounting for quantum fluctuations.

The standard treatment of Quantum Field Theory can also be applied to
gravitational fields
\cite{Gupta52,UtiyamadeWitt,Feynman63,Weinberg65,ZeldovichGrishchuk}.
It has however been early recognised that the infinities generated by radiative
processes due to gravitational fields cannot be dealt with usual
renormalisation
procedures \cite{t'HooftVeltman}.
A new theory of Quantum Gravity is needed in order to
describe all interactions including gravitation. Since such theory is not
yet available, discussions of the influence of quantum fluctuations on the
general principles associated with relativity have remained preliminary and
limited by approximations which remain uncontrolled by necessity.
In most approaches, Einstein's General Relativity is considered as a starting
basis
for a theory of gravitation which must remain valid at low energies. Due to
quantum fluctuations of gravitation, profound modifications are expected at
energies comparable to Planck energy, which should affect the ultimate
structure of space-time at very short distances \cite{Wheeler57}.
The interpretation of gravitation as geometric property of space-time,
which is inherited from the classical theory, has to find a new formulation
in quantum theory \cite{Rovelli91,Isham95}.
The question of the nature of a quantum vacuum including gravitation and,
hence, of a space-time being an arena for all interactions
still remains an open problem.
In the present paper, we shall try to provide hints for the construction of
such a
quantum framework, by emphasizing the necessity of a consistency between
vacuum field fluctuations and the property of relativity of motion.

\section{Relativity of motion in vacuum}

Modern mechanics is founded on the principle of inertia, that is the
conservation of uniform motion in the absence of external forces. This
property may equivalently be stated as relativity of uniform motion in empty
space. In particular rest is not a privileged state of motion. With the
advent of quantum theory, empty space is the place of vacuum fluctuations
and the symmetry of vacuum determines whether or not the principle of
relativity of motion remains valid for motion in quantum empty space.

In order to study the effect of motion on vacuum fields, one has first to
represent motion. Within the formalism of Quantum Field Theory, motion is
usually reduced to a mapping of space-time coordinates fitting
a classical trajectory in space-time. Arbitrary motion is represented by this
technique,
which however relies on the same classical approximation used when curved space
is
represented by a classical background metric \cite{BirrellDavies,Fulling}.
This approach perfectly fits the covariance techniques of General Relativity.
In the quantum regime however, it is known to generate paradoxes
due to a too crude mixing between quantum effects and classical descriptions of
motion \cite{deWitt62}.

A second approach puts emphasis on invariance properties in the spirit of
Special Theory of Relativity \cite{Einstein}. Since it is based upon symmetry
properties associated with space-time transformations, this approach is
limited to particular motions. As an emblematic example of this approach,
the principle of relativity of uniform motion is identified with the Lorentz
invariance of Maxwell equations \cite{EinsteinRelat}.
In Quantum Field Theory, the connection between vacuum fluctuations and
field propagation ensures that the electromagnetic vacuum also satisfies
invariance under Lorentz transformations. In the present section, we shall see
that the conformal invariance of electromagnetism allows to deal with
uniformly accelerated motion in the same manner as Lorentz invariance for
uniform motion \cite{JaekelReynaud95a,JaekelReynaud95b}.

As a first step,vacuum, which is the ground state, is characterised as the
particular case of thermal equilibrium state corresponding to a zero
temperature. For such a stationary state, different correlation functions
can be defined for any two fields $A$ and $B$
\begin{eqnarray}
C_{AB}(x) &=&<A(x)B(0)>-<A(x)><B(0)>=\hbar \sigma _{AB}(x)+\hbar \xi _{AB}(x)
\nonumber \\
2\hbar \sigma _{AB}(x) &=&C_{AB}(x)+C_{BA}(-x) \qquad \qquad 2\hbar \xi
_{AB}(x)=C_{AB}(x)-C_{BA}(-x)  \label{sacor}
\end{eqnarray}
The commutator $\xi _{AB}$ exhibits the quantum character of fields $A$ and $%
B$, while the anticommutator $\sigma _{AB}$ describes their fluctuations.
These functions are more easily manipulated in the Fourier domain. When using
the generic
definition
\begin{equation}
f(x)=\int {\frac{d^4k}{(2\pi )^4}}e^{-ik.x}f[k]
\end{equation}
$\xi _{AB}[k]$ is identified with a field propagator and $\sigma_{AB}[k]$
with a field noise spectrum. At thermal equilibrium, correlation
functions satisfy fluctuation-dissipation relations \cite{CallenWelton,Kubo66}
which are just a general expression of the Planck law \cite{Planck11}
\begin{equation}
C_{AB}[k]=\frac{2\hbar \xi _{AB}[k]}{1-e^{-\beta \omega }} \qquad \qquad
\sigma _{AB}[k]=\coth (\frac{\beta \omega }2)\xi _{AB}[k]   \label{Plancklaw}
\end{equation}
where $\omega $ is the field frequency $k_0$, $\beta =\frac \hbar {k_BT}$
characterises temperature $T$ while $\hbar $ and $k_B$ represent respectively
Planck and Boltzmann constants. For the sake of simplicity, the velocity of
light $c$ is taken as unit velocity. However, we have chosen to
write it explicitly in a few equations which may be used to evaluate orders
of magnitude. The zero temperature limit of equations (\ref{Plancklaw}) gives
relations
characteristic of vacuum ($\theta $ is Heaviside's step function and
$\epsilon $ is the sign function)
\begin{equation}
C_{AB}[k]=2\hbar \theta (\omega )\xi _{AB}[k]\qquad \qquad \sigma
_{AB}[k]=\epsilon (\omega )\xi _{AB}[k]  \label{fdv}
\end{equation}

{}For instance, vacuum fluctuations of the electromagnetic potential $A_\mu $
may be described by any one of the equivalent expressions
\begin{eqnarray}
C_{A_\mu A_\nu }[k] &=&2\hbar \theta (\omega )\xi _{A_\mu A_\nu }[k]
\qquad \qquad \xi _{A_\mu A_\nu }[k] = \pi \delta (k^2)\eta _{\mu \nu }
\nonumber \\
C_{A_\mu A_\nu }(x) &=&{\frac \hbar \pi }\eta _{\mu \nu }{\frac 1{%
(x_0-i\varepsilon )^2-{\bf x}^2}}  \label{vac}
\end{eqnarray}
where $\eta _{\mu \nu }$ is Minkowski metric, i.e. a diagonal matrix with
elements $(1, -1, -1, -1)$ on the diagonal, and
$\delta $ is Dirac's measure;
$x_0$ is the time delay and ${\bf x}^2$ the squared spatial distance between
the two points where fields are evaluated
($\varepsilon \rightarrow 0^{+}$, and Feynman gauge has been used).
Vacuum correlations of electromagnetic fields are thus characterised by a
spectrum limited to the light cone and to positive frequencies.
The spectrum is furthermore invariant under Lorentz transformations.
Conversely, the spectrum of vacuum field fluctuations
is completely determined by these properties, up to a constant
factor proportional to $\hbar $. Planck constant can be considered as fixing
the scale of vacuum or zero-point fluctuations \cite{Planck11,Boyer73} which
correspond to an energy of ${\frac 12}\hbar \omega $ per mode of frequency
$\omega $.

Invariance of vacuum under Lorentz transformations ensures that it has the
same appearance for all uniformly moving observers. The extension to
uniformly accelerated observers is confronted to the problem of finding a
representation for uniformly accelerated frames. As already noticed, there
exists a variety of possible representations which are all in agreement with
the same definition of uniformly accelerated motion \cite{Born09}. As a matter
of fact, vacuum fluctuations depend on the global definition of frame
transformations and not only on the definition of a single classical
trajectory. Infinitely many different changes of frame can be used to
transform a given accelerated trajectory to rest. A common choice within the
framework of General Relativity is to use Rindler coordinates \cite{Rindler}
which
are designed such that they transform rest into uniformly accelerated motion
while preserving the rigidity of bodies. Paradoxical properties arise when
discussing the transformation of quantum fields, since the sign of
frequencies is not preserved. This means that the number of particles and
even the distinction between particle and antiparticle is not
the same for different uniformly accelerated observers within this
formalism. In particular, vacuum is not invariant, but is transformed into a
thermal bath with a temperature proportional to the acceleration $a$
\cite{Unruh76}
\begin{equation}
\frac{\beta a}c=\frac{\hbar a}{k_BTc}=2\pi
\end{equation}

Paradoxes and conflicting views are raised by this result.
On one hand, the detection of thermal quanta by a detector moving in
vacuum with uniform acceleration has been predicted, using a simplified
model of detector. However, the detection process in the moving frame has to
be interpreted differently by an observer at rest, for whom the excitation
of the detector is accompanied by the emission of a quantum.
This leads to conflicts with the principle of energy conservation, the
solution of which require subtle reasoning \cite{UnruhWald,Unruh92}. A few
suggestions have been made for the observation of the modification of vacuum
by acceleration \cite{BellLeinaas,Rogers88,Rosu94}. On the other hand,
it has also been
argued that an oscillator moving with uniform acceleration in vacuum does
not radiate \cite{Grove} neither detects thermal quanta \cite
{RaineSciamaGrove}. The paradoxes raised by detection have also been
discussed by modelling the detector as a two level atom \cite{AudretschMuller}.
The consequences for the equivalence principle of the transformation of
vacuum into a thermal bath have been discussed, leading to the
introduction of different vacua depending on the physical situation and the
observer \cite{GinzburgFrolov}.
In view of these difficulties, two simple points may be recalled here.
First, Rindler transformations do not preserve Maxwell equations.
In other words, the field propagator is changed under such a transformation.
In these conditions, it is not so surprising to obtain a change of the vacuum
spectrum or of the relation between the vacuum spectrum and the propagator.
Then, a thermal bath has quite different physical properties than vacuum.
It leads to a friction force for uniform motion so that
uniform velocity is no longer equivalent to rest in a Rindler frame. But
there is no simple composition rule for different Rindler transformations or
even for Rindler transformations and Lorentz transformations.
Rindler transformations do not form a group.

Another representation of uniformly accelerated frames can be used
which is more adapted to the point of view of symmetry properties.
Shortly after Einstein built up the Special Theory of Relativity, Bateman
and Cunningham \cite{Bateman09,Cunningham09} established that Maxwell equations
possess a symmetry group larger than the Poincar\'{e} group of
translations and Lorentz transformations. This group contains conformal
coordinate transformations defined as those transformations for which the
modified metric keeps the form of Minkowski metric, up to a conformal
factor. The group may be generated by Poincar\'{e} transformations and the
transformations of space-time coordinates $x^\mu$ which can be
built with translations and four-dimensional inversion
and which are defined by
\begin{equation}
x^\mu \rightarrow {\bar{x}}^\mu, \qquad
{\frac{{\bar{x}}^\mu }{{\bar{x}}^2}}={\frac{x^\mu }{x^2}}-a^\mu  \label{conf}
\end{equation}
where $ a^\mu $ plays the role of an acceleration four-vector.
These transformations may be identified as transformations to accelerated
frames. To be more precise, uniformly accelerated motion may be defined
in Relativity Theory by the condition that the Abraham vector vanishes
\cite{Born09}. This vector $\Gamma_{\mu}$ is defined, for any trajectory
$x_{\mu}(\tau)$, as ($\tau$ is a proper time defined along the trajectory)
\begin{equation}
\Gamma^{\mu}={\frac{d^3x^\mu }{d\tau ^3}}
+({\frac{d^2x}{d\tau ^2}})^2{\frac{dx^\mu }{d\tau }} \qquad \qquad
d\tau = \sqrt{\eta _{\mu \nu }dx^\mu dx^\nu}  \label{abv}
\end{equation}
Conformal transformations preserve the Abraham vector (\ref{abv})
up to a factor \cite{Hill45}. Hence, they also preserve
the definition of uniformly accelerated motion. This means that rest is
transformed
into uniformly accelerated motion under a conformal transformation
and, conversely, that a conformal transformation can be
found which transforms a given uniformly accelerated motion into rest. For
that reason, it has been possible to use conformal transformations \cite
{Page36,Gupta61,FultonRohrlichWitten} to discuss the relativistic effects of
uniform acceleration \cite{EinsteinAccel}.

Since Maxwell equations are invariant under conformal transformations, the
propagator of electromagnetic fields is also preserved and it is worth
studying the transformation of vacuum correlation functions. It has in fact
been demonstrated that these functions are also invariant \cite
{JaekelReynaud95a}. The noise spectrum
$\sigma _{F_{\mu \nu}F_{\rho \sigma }}$ which characterises the fluctuations
of the electromagnetic field
\begin{equation}
F_{\mu \nu }=\partial _\mu A_\nu -\partial _\nu A_\mu
\end{equation}
has exactly the same form in the accelerated frame when written in terms of
the transformed coordinates as in the inertial frame when written in terms
of the original coordinates.
The noise spectrum $\sigma
_{A_\mu A_\nu }$ which characterises the fluctuations of the electromagnetic
potential has its form changed but this change may be eliminated by
gauge transformation. Similarly, Maxwell equations written in terms
of the electromagnetic potential are invariant up to a gauge
transformation \cite{Dirac36,Bhabha36}.

We may emphasize that correlation functions are written in
the conformal accelerated frame as well as in the inertial frame
in terms of a Minkowski metric.
Hence, the preservation of vacuum fluctuations is not a covariance
statement which would merely mean that the change of form due to a
coordinate transformation is compensated by the corresponding
change of the metric. This preservation should rather be
interpreted as an invariance property under conformal transformations,
which generalises the known invariance property
under Poincar\'e transformations.

In order to emphasize the symmetry properties associated with conformal
invariance, it is worth introducing the conformal algebra
\cite{ItzyksonDrouffe} associated with the group formed by conformal coordinate
transformations. The generators of this algebra include Lorentz generators
$J_{\mu \nu }$ and translations ($P_\mu $) which form the Poincar\'{e} algebra
\begin{eqnarray}
&&\left( P_\mu ,P_\nu \right) =0 \qquad \qquad \left( J_{\mu \nu },P_\rho
\right
)
=\eta _{\nu \rho }P_\mu -\eta _{\mu \rho }P_\nu   \nonumber \\
&&\left( J_{\mu \nu },J_{\rho \sigma }\right) =\eta _{\nu \rho }J_{\mu
\sigma }+\eta _{\mu \sigma }J_{\nu \rho }-\eta _{\mu \rho }J_{\nu \sigma
}-\eta _{\nu \sigma }J_{\mu \rho }
\end{eqnarray}
as well as a dilatation generator ($D$) and four generators ($C_\mu $) for
transformations to accelerated frames (\ref{conf}) satisfying the following
commutation rules
\begin{eqnarray}
&&\left( D,P_\mu \right) =P_\mu \qquad \qquad \left( D,J_{\mu \nu }\right) =0
\nonumber \\
&&\left( P_\mu ,C_\nu \right) =-2\eta _{\mu \nu }D-2J_{\mu \nu }  \nonumber
\\
&&\left( J_{\mu \nu },C_\rho \right) =\eta _{\nu \rho }C_\mu -\eta _{\mu
\rho }C_\nu   \nonumber \\
&&\left( D,C_\mu \right) =-C_\mu \qquad \qquad \left( C_\mu ,C_\nu \right) =0
\label{ConfAlg}
\end{eqnarray}
The total number $N$ of photons is invariant under all these
conformal transformations
\begin{eqnarray}
\left( P_\mu ,N\right) =\left( J_{\mu \nu },N\right) =\left( D,N\right)
=\left( C_\mu ,N\right) =0
\end{eqnarray}
As a particular case, the vacuum state defined by $N=0$ is invariant under
conformal transformations. More generally, the notion of $N$ photon states
is the same for inertial or uniformly accelerated observers. This allows to
discuss how the frequency of a photon is changed in a transformation to
accelerated frames and, hence, to establish a connection between Einstein
equivalence principle and the domain of quantum fluctuations \cite
{JaekelReynaud95b}.

Conformal invariance allows a simple
discussion of scattering of vacuum fields by uniformly accelerated mirrors.
A scatterer uniformly accelerated in vacuum may indeed be transformed into a
scatterer at rest through a conformal transformation which leaves vacuum
invariant. In other words, vacuum as defined for an inertial observer
remains vacuum for a uniformly accelerated observer. As an immediate
consequence, a mirror moving with uniform acceleration in vacuum behaves
like a mirror at rest and does not radiate. The absence of radiation may be
stated
as the preservation of relativity of motion for
uniformly accelerated motion in vacuum \cite{JaekelReynaud95a}.

This result has in fact been obtained in a number of specific cases by
explicit computation. Fulling and Davies were the first ones to establish that
the
force experienced by a perfect mirror moving in a two-dimensional scalar
vacuum exactly vanishes for uniformly accelerated motion \cite{FullingDavies}.
This result was extended to the case of plane mirrors moving in
four-dimensional
space-time, in conformally invariant scalar fields \cite{FordVilenkin} or
electromagnetic fields \cite{MaiaNeto94}. For all these calculations, the free
quantum field theory obeys conformal invariance and the scattering upon the
perfect mirror also obeys conformal invariance. The general result reported
in the present section has a larger domain of validity since the second
condition does not appear to be necessary. Actually, the absence of radiation
for uniformly accelerated motion has also been established for scatterers
which do not obey conformal invariance like partly transmitting mirrors
where a frequency scale is given by the reflection cutoff \cite
{JaekelReynaud92} and Fabry-Perot cavities where a length scale is given by
the distance of the two mirrors \cite{JaekelReynaud93b}.

\section{Dissipative motion in vacuum}

The presence of quantum field fluctuations implies that scatterers are
submitted to a fluctuating radiation pressure, even when they are at rest in
vacuum \cite{Barton91}. Preservation of inertial motion in
vacuum corresponds to the conservation of the mean values of energy and
momentum but it is also a consequence of conservation laws that a scatterer
at rest in vacuum feels a fluctuating force \cite{JaekelReynaud92}.
These force fluctuations and
the associated dissipation are analysed in detail in this section, which
begins by discussing the fluctuations of the vacuum stress
tensor which determine the radiation pressure exerted by electromagnetic
fields upon scatterers.

Maxwell stress tensor $T_{\mu \nu }$ has vacuum correlations which may be
determined from those of the electromagnetic field (\ref{vac})
\begin{eqnarray}
C_{T_{\mu \nu }T_{\rho \sigma }}[k] &=&{\frac{\hbar ^2}{40\pi }}\theta
(\omega )\theta (k^2)(k^2)^2\pi _{\mu \nu \rho \sigma }  \nonumber \\
\pi _{\mu \nu \rho \sigma } = {\frac 12}(\pi _{\mu \rho }\pi _{\nu \sigma
}+\pi _{\mu \sigma }\pi _{\nu \rho })&-&{\frac 13}\pi _{\mu \nu }\pi _{\rho
\sigma }  \qquad \qquad
\pi _{\mu \nu } = \eta _{\mu \nu }-\frac{k_\mu k_\nu }{k^2}  \label{CTT}
\end{eqnarray}
Like for vacuum correlations of fields, the spectra are completely
determined in vacuum by their symmetry properties, up to a numerical factor.
Indeed, stress tensors are symmetric in their two indices, while conformal
symmetry results in a vanishing trace. Energy-momentum conservation implies
that divergence of the stress tensor vanishes, so that its tensorial form is
completely
determined. Lorentz invariance limits energy-momentum dependence to Lorentz
scalars, and to the interior of the light cone and positive frequencies.
{}Finally, fluctuations of quadratic forms in the fields must scale like $\hbar
^2$.
Radiation pressure fluctuations thus take a form which is
determined by the symmetries due to the conservation laws and the
constraints associated with vacuum \cite{Barton91,JaekelReynaud95c}.

The fact that a scatterer is submitted to a fluctuating radiation pressure in
vacuum,
even when it stays at rest, signals the existence of forces opposing to motion
of the scatterer.
As emphasized in the Introduction, the existence of vacuum fluctuations
drastically affects the problem of motion in empty space. When a scatterer
is set into motion, the quantum fields which are scattered are perturbed and
this perturbation modifies energy-momentum balance. Energy-momentum is
exchanged between field and scatterer, leading to a reaction force exerted
by vacuum fields on the mirror. The radiation reaction force has first been
studied within the formalism of Quantum Field Theory by considering
mirrors as perfectly reflecting boundaries for the quantum field
\cite{deWitt75}.
In the case of a scalar field scattered by a perfect mirror in a
two-dimensional space-time, the radiation reaction force has been
shown to be proportional to the Abraham vector (\ref{abv})
for any mirror's trajectory \cite{FullingDavies}
\begin{equation}
{}F^{\mu } = {\hbar \over 6 \pi c^2} \Gamma^{\mu }  \label{FD}
\end{equation}
Radiation reaction vanishes for uniform acceleration, due to conformal symmetry
of vacuum. This expression also shows that non uniform acceleration
is accompanied by radiation. This approach may be applied
to arbitrary motions, but it is limited to perfect mirrors and is plagued
with infiniteness problems well-known from the study of Casimir energy.

In early derivations \cite{Casimir48}, Casimir energy was interpreted as a
modification of field eigenfrequencies, and hence of the zero-point energy,
due to the presence of boundaries. Once two reflectors are present, vacuum
energy depends on their distance, giving rise to mean Casimir forces. This
approach had to face the infiniteness of vacuum energy and formal
prescriptions were introduced to obtain finite results \cite{BirrellDavies}.
It was however suggested that a more satisfactory
solution would be to suppose mirrors to be transparent at the limit of high
frequencies. This is certainly a reasonable assumption for any realistic
mirror which, furthermore, eliminates the infiniteness problem. After
Casimir forces were given a local interpretation as due to the radiation
pressure of vacuum fields \cite{BrownMaclay}, a scattering matrix approach
was developed to deal with frequency-dependent reflectivities. This
approach allowed to obtain finite forces without encountering any infinite
quantity \cite{JaekelReynaud91}. It has also been used to study the case of
narrow-band dielectric mirrors and to compute Casimir forces limited by the
mirrors' scattering bandwidths \cite{Iacopini93,LambrechtJaekelReynaud97}.
This approach was then extended to the case of moving mirrors by making use
of the linear response formalism \cite{JaekelReynaud92}. It is thus
restricted to motions of small amplitude, but it allows to show that the
force induced in reaction to motion is related to the force fluctuations
felt by mirrors at rest through general fluctuation-dissipation relations.

Relations between fluctuations and dissipation were first established by
Einstein, in the case of Brownian motion in a thermal bath \cite
{EinsteinBrownian}. From energy-momentum conservation and thermal
equilibrium, Einstein showed that the Brownian diffusion of momentum is
accompanied by a dissipative force opposed to the scatterer's motion and
characterised by a friction coefficient depending on the diffusion
coefficient and the bath temperature. Fluctuation-dissipation relations
were first given their full quantum form by Callen and Welton \cite
{CallenWelton} and later generalised to the linear response formalism by
Kubo \cite{Kubo66}.
The radiation reaction force can be seen to emerge quite generally as a
particular kind of causal response function. For motions of small
amplitude, the linear response formalism can be applied and the derivation
of the reaction force can be briefly sketched as follows. Whatever the
Hamiltonian of the total system is, it can be decomposed into free and
interacting parts
\begin{equation}
H=H_0+H_I  \qquad \qquad
\delta H_I=-\delta q(t)F(t)  \label{mop}
\end{equation}
The latter part describes the interaction between the field
and the sources on the mirror, which is perturbed
by the motion of the scatterer. At first order,
this perturbation is proportional to the
scatterer's displacement $\delta q$ and its generator
is identified as the force $F$ felt by the scatterer \cite{JaekelReynaud92}.
This relation can also be seen as the result of transformations
under displacements generated by the field momentum, provided
the interaction is translation invariant, i.e. the total energy and momentum
are
conserved.

The linear response of any function $A$ of the interacting fields
to the perturbation is then described by a linear susceptibility which is
directly related to the commutator of this function with the generator
\begin{eqnarray}
<\delta A(t)>=\int_{-\infty }^\infty dt^{\prime }\chi _{AF}(t-t^{\prime
})\delta q(t^{\prime })  \qquad &\qquad&
<\delta A[\omega ] > = \chi_{AF} [\omega]  \delta q[\omega ] \nonumber \\
Im(\chi _{AF}) [\omega ] &=& \xi _{AF}[\omega ]  \label{fd}
\end{eqnarray}
In other words, the dissipative part of the susceptibility coincides with
the commutator spectrum.
The reactive part of the susceptibility may be reconstructed
from the dissipative part through dispersion relations, but short time
singularities might require that subtractions are accounted for in
this reconstruction \cite{Barton63}. Fluctuation-dissipation relations (\ref
{fd}) characterise the linear response of a quantum system to a perturbation
\cite{Kubo66}. At thermal equilibrium, the commutator spectrum $\xi _{AB}$ is
directly related to correlation functions \cite{CallenWelton}. At zero
temperature in particular, the relations (\ref{fdv}) characteristic of
vacuum fluctuations are obtained.

Clearly, the motional susceptibility of the force itself is related to the
fluctuations of the force exerted on the scatterer at rest, according to
(\ref{fd})
\begin{eqnarray}
&&<\delta F[\omega ] > = \chi _{FF}[\omega ] \delta q [\omega ] \nonumber \\
&&C_{FF} [\omega ] = 2\hbar \theta (\omega ) \xi _{FF}[\omega ]
= 2\hbar \theta (\omega ) Im(\chi _{FF}[\omega ])
\end{eqnarray}

To illustrate these properties, let us consider the simple case of a
perfect mirror scattering a scalar field in a two-dimensional space-time. In
that case, stress tensor fluctuations (\ref{CTT}) lead to the following
fluctuations for the momentum density $p$ of incoming
fields at an arbitrary point
\begin{equation}
C_{pp}[\omega ]={\frac{\hbar ^2}{12\pi }}\theta (\omega )\omega ^3
\end{equation}
{}For a perfect mirror, all incoming momentum is reflected back and, hence, the
radiation pressure is twice the incoming momentum density
\begin{equation}
C_{FF}[\omega ]={\frac{\hbar ^2}{3\pi c^2}}\theta (\omega )\omega ^3
\end{equation}
Hence, a force proportional to the third time
derivative of the mirror's position is obtained \cite{JaekelReynaud92}
\begin{equation}
\xi _{FF}[\omega ] ={\frac \hbar {6\pi c^2}}\omega ^3  \qquad \qquad
<\delta F(t)> ={\frac \hbar {6\pi c^2}}\delta q^{\prime \prime \prime }(t)
\label{qtierce}
\end{equation}
This is the dissipative force (\ref{FD}) for a perfect mirror moving in
vacuum \cite{FullingDavies} at the limit where the velocity remains
much smaller than the velocity of light.

The dissipative force felt by perfect mirrors moving in four-dimensional
space-time and in vacuum of different quantum fields have similarly been
computed and shown to involve higher derivatives of the mirror's position
\cite{FordVilenkin,MaiaNeto94,BartonEberlein92}. As in the two-dimensional
case,
they are connected through fluctuation-dissipation relations with force
fluctuations evaluated for three-dimensional mirrors which could have
different shapes \cite{Barton91,Eberlein92,MaiaNetoReynaud}. This connection
has also been established for mirrors described by frequency-dependent
scattering matrices \cite{JaekelReynaud92}.
At this point, it is worth discussing the analogy with the problem of
radiation reaction in ElectroDynamics.
It has been known for long \cite{Pauli} that a charge moving in vacuum with
non uniform acceleration radiates and that the loss of energy-momentum
gives rise to the Abraham-Lorentz radiation reaction force
which is proportional to Abraham vector (\ref{abv}).
The equation of motion for a charge in vacuum has also been written as a
Quantum Langevin equation, with the radiation reaction force and the
fluctuating Langevin force satisfying fluctuation-dissipation relations
\cite{Dekker85,FordLewisOConnell}, in close analogy with relations
(\ref{qtierce}). The fact that electric charges and dipoles feel a
fluctuating vacuum field, the properties of which are characterized by
a noise spectrum proportional to $\omega ^3$, has been thoroughly
demonstrated by studies of spontaneous emission processes
\cite{Cohen,Milonni}. It is also known from
discussions of classical electron theory, that an equation of motion
involving third derivatives leads to instabilities or violations of
causality \cite{Rohrlich}. In the case of mirrors, this difficulty
points to inconsistencies of the
model of perfect mirror which can be circumvented by considering
mirrors transparent at field frequencies higher than a cut-off frequency.
Under the condition that the energy corresponding to the cutoff
is smaller than the mass of the mirror, the causal scattering of the field and
the
characteristic properties of vacuum spectra lead to a
mirror's susceptibility which is itself a passive,
and hence causal, function \cite{JaekelReynaud92b}.
This condition for causal motions is violated by perfect mirrors.
In fact, the renormalisation procedure used for perfect mirrors,
as well as for electrons, to render finite an infinite mass,
is incompatible with a causal mechanical description \cite{Dekker85}.

The modelisation of mirrors as point-like structures does not give account
of important physical consequences of spatial extension.
Cavities built with two mirrors appear as the simplest systems to discuss
such consequences.
A lot of discussions has been devoted to the production of radiation inside
a cavity built with two perfectly reflecting mirrors
\cite{Dodonov,SassaroliSrivastava}.
However, these calculations do not
consider the photons radiated by the cavity which is treated
like a closed system. Even for the photons produced inside the cavity,
they disregard the important problem of
finite lifetime of photons inside the cavity.
When built with partly transmitting mirrors in contrast,
the cavity appears as an open system able to radiate into the free
field vacuum.
Moreover, exploiting the finesse of the cavity to
amplify the emitted radiation may bring dissipation of motion in vacuum
within the realm of experimental observation
\cite{BraginskyKhalili91,LambrechtJaekelReynaud96}.
Two mirrors in vacuum not only feel the static Casimir force, but also
fluctuating forces. They
also undergo dissipative forces when they are set into motion,
since the fields scattered by the two mirrors are affected by their motion.
As in the case of a single mirror, the motional Casimir forces $F_i$
can be obtained at first order in the mirrors' displacements $\delta q_j$
in terms of linear susceptibilities \cite{JaekelReynaud93a}
\begin{equation}
<\delta F_i[\omega ]>=\sum_j\chi _{F_iF_j}[\omega ]\delta q_j[\omega ]
\label{chiij}
\end{equation}
The motional susceptibilities are related through
fluctuation-dissipation relations to the fluctuations of Casimir forces
felt by the mirrors at rest.

The mechanical susceptibilities can be first studied for very slow
motions, i.e. in the quasistatic limit where the motional susceptibilitites
are dominated by reactive contributions. At zero frequency, these contributions
include the variation of the Casimir force with the mirror's distance.
For low frequencies, they also contain terms which correspond to
modifications of the inertial responses of the mirrors. These inertial
forces show remarkable relativistic properties. They include not only mass
corrections for each mirror, depending on the mirrors' distance, but also a
force acting on a mirror and depending on the other's acceleration. When
taken altogether and for a global motion of the cavity, the resulting
inertial correction corresponds to the inertia of the Casimir energy of the
cavity.
Hence, inertial forces emerge for a cavity with finite spatial extension,
although no radiation is predicted,
in consistency with the conformal invariance of vacuum.
This also shows that the law of inertia of energy still holds for
energies due to vacuum fluctuations \cite{JaekelReynaud93b}. The energy due
to the Casimir forces can furthermore be seen as an energy stored inside the
cavity, as a result of the scattering time delays undergone by field
fluctuations. Like the radiation pressure, the field energy inside the
cavity fluctuates. As a consequence, the inertial mass of the cavity in
vacuum has its proper quantum fluctuations \cite{JaekelReynaud93c}.

Quite generally, the susceptibilities (\ref{chiij}) involve the same Airy
function
which determines the energy spectral density inside the cavity \cite{Haroche}.
As a result, motional Casimir forces are resonantly enhanced when
the mechanical frequency $\omega$ associated with motion is an integer
multiple of optical resonance frequencies of the cavity
\cite{LambrechtJaekelReynaud96}
\begin{equation}
\omega=n{\frac \pi \tau}  \qquad \qquad n \geq 2  \label{omegarad}
\end{equation}
$\tau$ is the field propagation time from one mirror to the other.
At resonance, the displacements of the mirrors also induce an
important redistribution of the energy density inside the cavity,
leading to the formation of energy peaks \cite{Law94,ColeShieve}.
As for a single mirror, the motions of the mirrors of the cavity
induce a radiation, but the latter is now enhanced for resonant motions.
Let us consider the case of two mirrors following harmonic
motions at a frequency $\omega $ and with an amplitude $a$.
For a long oscillation time $T$ and a cavity with a high finesse ${\cal F}$,
the number $N$ of radiated photons is found to be proportional to
the product of three dimensionless numbers
\begin{equation}
N \simeq \frac{\omega T}{2\pi }\frac{v^2}{c^2} {\cal F}  \qquad \qquad
v \simeq \omega a    \label{Nrad}
\end{equation}
The first factor is the number of mechanical oscillation periods during
the time $T$ and the second one is the square of
the peak velocity  $v/c$ of the vibrating mirrors.
These two first factors represent the result expected for a single
vibrating mirror, and the corresponding order of magnitude is
vanishingly small for any reasonable material velocity.
The third factor represents a resonance enhancement proportional to
the finesse and allows to hope
that dissipation of motion in vacuum may be brought
within the realm of experimental observation, provided very high
finesses are used \cite{LambrechtJaekelReynaud96}.
In equation (\ref{Nrad}), the mechanical oscillation frequency has been
supposed to be equal to one of the resonance frequencies (\ref{omegarad}).
Precisely, even modes $\omega =\frac{2\pi }\tau ,\frac{4\pi }\tau \ldots $
correspond to elongation modes with a
periodic modulation of the mechanical cavity length
while odd modes $\omega =\frac{3\pi }\tau ,\frac{5\pi }\tau \ldots $
are excited by a global translation of the cavity with its length kept
constant.
{}For the latter effect as well as in the case of a single oscillating mirror,
the cavity moves in vacuum without any further reference than vacuum
itself. For the cavity however, radiation is now enhanced by the cavity
finesse.
We may emphasize that the number of radiated photons diverges at the limit of
perfectly reflecting mirrors where the finesse goes to infinity.
This shows once more that the simple model which represents mirrors as
perfect reflectors and, hence, the cavity as a closed system
misses important physical phenomena.

\section{Vacuum fluctuations and localisation}

The existence of quantum fluctuations leads to reconsider the notion of
position and, hence, of motion which can be considered as a set of different
positions in time. Quantum Mechanics has early been known to modify the
concept of position to account for the incompatibility between a definite
position and a definite velocity. Heisenberg inequalities constrain position
and velocity uncertainties in a given quantum state and connect them
to the commutator between position and momentum.
These inequalities also provide a standard quantum limit for measurements
of position \cite{Caves,Yuen,Ozawa}.

The sensitivity limits are more precisely analysed
in terms of noise spectra, which describe the quantum fluctuations
associated with non commuting variables \cite{JaekelReynaud90}. This technique
has been thoroughly used in Quantum Optics \cite{ReynaudHeidmann}.
It also allows to discuss quantum fluctuations of position within the
framework of Quantum Field Theory. As a consequence of quantum
force fluctuations, scatterers in vacuum undergo a quantum Brownian motion
\cite{JaekelReynaud93d}. Positions have fluctuations related to
force fluctuations and, in particular, they inherit their intrinsic
quantum character even if they are {\it a priori} treated as classical
variables. Considered from the point of view of Quantum Measurement Theory,
position is reached by means of quantum fields used as probes in the
measurement process \cite{deWitt63}.
Then quantum limits also emerge as a consequence of
quantum field fluctuations. It is a remarkable consistency result that these
two approaches agree on the ultimate limits constraining localisation in
vacuum.

The quantum Brownian motion induced on position by the force fluctuations in
quantum vacuum has been studied within the framework of linear response
formalism \cite{JaekelReynaud93d}. For small displacements,
the effects of motion on field scattering and of field radiation pressure
on the scatterer's motion can be treated as reciprocal linear responses
\begin{eqnarray}
{}F[\omega ] &=&\chi _{FF}[\omega ]q[\omega ]+F^{in}[\omega ]  \nonumber \\
q[\omega ] &=&{\frac 1{m(\omega _0^2-(\omega +i\epsilon )^2)}}
F[\omega]+q^{in}[\omega ]  \label{lin}
\end{eqnarray}
The mirror has been assumed to be anchored elastically with a proper
frequency $\omega _0$.
The solution of the linear system (\ref{lin}) provides the position
and radiation pressure of the coupled system in terms of input fluctuations.
In fact, input position fluctuations $q^{in}$, which
are localised at the proper frequencies $\pm \omega _0$ of the free
oscillator, do not feed the final fluctuations in the coupled system, which
are thus only determined by input force fluctuations $F^{in}$. Fluctuations
of positions in the coupled system are then correctly described by a quantum
Langevin equation, which includes input force fluctuations $F^{in}$ and the
mirror's mechanical response function $\chi _{qq}$
\begin{equation}
\chi _{qq}[\omega ]={\frac 1{m_0(\omega _0^2-\omega ^2)-\chi _{FF}[\omega ]}}
\label{mrf}
\end{equation}

The coupled system also obeys fluctuation-dissipation relations.
In particular, the noise spectrum $C_{qq}$ which characterizes position
fluctuations and the susceptibility $\chi _{qq}$ which describes
mechanical response of the mirror to an input force are directly connected
to each other
\begin{equation}
C_{qq}[\omega ] = 2\hbar \theta (\omega ) \xi _{qq}[\omega ]
= 2\hbar \theta (\omega ) Im(\chi _{qq}[\omega ])  \label{qfp}
\end{equation}
The noise spectrum $C_{qq}$ contains resonance peaks which correspond
to the quantum fluctuations of position associated with Schr\"{o}dinger's
propagator for the free oscillator.
In fact, these fluctuations may be seen as those fluctuations which remain
in the non dissipative limit of decoupling, for a scatterer coupled to the
radiation pressure fluctuations of vacuum quantum fields.
Besides the resonance peaks associated with Schr\"{o}dinger's propagator,
the noise spectrum $C_{qq}$ includes a background which spreads over all
frequencies and describes the noise added by vacuum pressure fluctuations on
position
fluctuations. Considering the particular case of an unbound mirror
($\omega_0=0$),
the background may be estimated from (\ref{qfp}) as
\begin{equation}
C_{qq}[\omega ]\simeq \lambda _c^2{\frac{\theta (\omega )}\omega }\qquad
\qquad \lambda _C={\frac \hbar {mc}}   \label{cqqev}
\end{equation}
where $\lambda _C$ is the Compton wave-length associated with mass $m$.
As a consequence, an unbound mirror is shown to undergo
a quantum diffusion behaviour with an order of magnitude
given by the Compton wavelength  \cite{JaekelReynaud93d}.

The quantum nature of position can also be investigated from
the point of view of Quantum Measurement Theory.
Canonical commutation relations suggest that
independent measurements performed at different points in space-time are
constrained by a standard quantum limit \cite{Caves,Yuen,Ozawa}. Similar
constraints apply to field measurements, as early discussed by Bohr and
Rosenfeld \cite{BohrRosenfeld}. The motion of the test charge used to
measure the field strength results in a radiated field which perturbs a
further measurement. The perturbation is given by the field propagator, that is
also
by the field commutator, between the two measurement points.
Motion and field measurements are limited by the backaction of
the probe field or test particle used, expressed by a response function or a
commutator  \cite{deWitt63}.

High sensitivity measurements of position have been reanalysed recently \cite
{Braginsky} and more emphasis has been put on the role played by the
measurement
strategy in fixing ultimate limitations \cite{BraginskyKhalili,BockoOnofrio}.
The standard quantum limit does not correspond to an optimal strategy, since
it corresponds to independent successive measurements. A better exploitation
of quantum correlations allows to attain lower limits
\cite{JaekelReynaud90,Unruh83,Luis92,Pace93}.

This is illustrated by the example of an interferometric measurement of
position,
which is a prototype of a space-time measurement of high precision
\cite{Caves,Grangier87,MinXiao87}.
In such a measurement, the phase $\varphi $ of a probe field is monitored.
An estimation of the mirror's position $q$
is deduced ($K_0$ is the mean wave-vector of the incident field)
\begin{equation}
q = \frac{\varphi}{2K_0}  \label{pf}
\end{equation}
Hence, the inferred position is directly affected by phase fluctuations $\delta
\varphi $
of the probe field. Simultaneously, each reflected photon of the probe field
transmits twice its momentum to the mirror, so that fluctuations $\delta I$
of the photon flux $I$ in the probe field result in a fluctuating radiation
pressure
\begin{equation}
\delta F=2\hbar K_0 \delta I  \label{if}
\end{equation}
The mirror moves under these force fluctuations and the measured position is
affected,
which reveals a simple example of backaction during measurement.
{}Finally, the measurement adds a noise $\delta q$ on the estimated position
which is obtained as
a linear combination of conjugate input field fluctuations (\ref{pf}-\ref{if})
\begin{equation}
\delta q={\frac {\delta \varphi} {2K_0}} + 2\hbar K_0 \chi _{qq} \delta I
\qquad \qquad
\chi _{qq}={\frac 1{m(\omega _0^2-\omega ^2-i\gamma \omega )}}
\end{equation}
$\chi _{qq}$ is the mechanical susceptibility of the mirror to an applied
force, assuming that it is elastically anchored at frequency
$\omega _0$ with damping mechanisms gathered in a coefficient $\gamma $.

The phase fluctuations $\delta \varphi $ and intensity fluctuations $\delta I$
obey generalised Heisenberg inequalities at each frequency.
As a result, the noise spectra associated with $\delta q$ take minimal values
which fix lower bounds on the measurement sensitivity.
When the probe fields have uncorrelated phase and intensity fluctuations,
the lower bound is essentially determined by the modulus of the mechanical
susceptibility
\begin{equation}
\sigma _{\delta q \delta q}[\omega ] \ge |\chi_{qq}[\omega ]|  \label{sql}
\end{equation}
This is the standard quantum limit which may be
also expressed as a spectral density for the noise energy equal to
Planck constant $\hbar$.
A much better sensitivity is obtained \cite{JaekelReynaud90}
by using optimally squeezed fields chosen such as to minimise
the fluctuations of the particular combination of the two quadrature
components entering the output noise $\delta q$.
Measurement sensitivity is thus limited by the dissipative part of the
mechanical susceptibility
\begin{equation}
\sigma _{\delta q \delta q}[\omega ] \ge |Im\chi_{qq}[\omega ]|  \label{uql}
\end{equation}
At the limit of a small damping ($\gamma \ll \omega $), the dissipative part is
much
lower than the reactive part and the ultimate quantum limit (\ref{uql}) lies
far beyond
the standard quantum limit (\ref{sql}).

{}Fluctuations of position result from the quantum Brownian motion
induced by vacuum radiation pressure fluctuations, which are also related
(see (\ref{qfp})) to the dissipative part of the mirror's mechanical
susceptibility
(\ref{mrf}). In a position measurement, the final noise will include
permanent fluctuations induced by radiation pressure of vacuum fields
and fluctuations added by the probe field.
An optimal measurement will have its
sensitivity limited by the dissipative part of the mechanical susceptibility.
This dissipative part contains an
irreducible component corresponding to vacuum fluctuations
which sets the ultimate quantum level of sensitivity
in a position measurement.

\section{Gravitational vacuum and space-time}

This ultimate noise associated with vacuum radiation pressure has an
order of magnitude mainly determined by the Compton wave-length
$\lambda _C$ associated with the mass $m$ of the scatterer (see
eq.(\ref{cqqev})).
When the mass $m$ is greater than Planck mass $m_{P}$,
this limit on position measurement goes beyond
the Planck length $l_{P}$
\begin{equation}
m_{P} = \sqrt{\hbar c \over G} \sim 22\mu\mbox{g}
\qquad \qquad
l_{P} = \sqrt{\hbar G \over c^3} \sim 1.6 10^{-35}\mbox{m}
\end{equation}
where $G$ is the Newton constant.
But sensitivity in length measurement is expected to be limited by Planck
length
in any sensible quantum theory \cite{Wheeler57,deWitt62b,Padmanabhan87}.
This implies that, at such a level of sensitivity, quantum fluctuations of
gravitation
must be taken into account when discussing ultimate quantum limits
\cite{JaekelReynaud94}.
It is also known since Einstein \cite{EinsteinGR} that metrical properties of
space-time
depend on the presence of a gravitational field. Localisation in space-time has
thus
to be affected by gravity and, particularly, by quantum fluctuations of
gravity.
It follows that the very notion of localisation in vacuum has to be
reconsidered in view of its relation with gravitation.

Since the discovery of vacuum fluctuations, the question of their
gravitational contribution has been a matter of debate. Problems with the
infinite energy of vacuum fluctuations early led to the view that
vacuum energy should not contribute to gravity. It was noticed that
vacuum energy may be forced to vanish by definition
\cite{Pauli,FeynmanHibbs,Enz74}.
This is done by prescribing that any expression involving
quantum fields has to be normally ordered with creation
and annihilation operators arranged so that the expression
vanishes in vacuum. This normal ordering prescription
relies on a distinction between positive and negative field frequencies
which depends on the choice of coordinate map.
It is therefore incompatible with the covariant description required by
General Relativity.
These problems are particularly acute in the context of cosmology
\cite{Zeldovich81,Weinberg89,Wesson91,AdlerCasey}.
They cannot be solved by using a classical description of curved space
as a background for quantum fields \cite{deWitt62}. Among the difficulties
which arise in this domain, lies the need to regularise the infinite
energy-momentum
tensor \cite{BirrellDavies,Fulling}. This procedure gives rise to ambiguities
and to anomalies, that is to a breakdown
at the quantum level of usual symmetry properties of the energy-momentum
tensor \cite{Christensen76,Dowker77,Wald78,Horowitz80}.

On the other hand, it is well known that gravitational fields can be dealt
with by using standard techniques of Quantum Field Theory
\cite{Feynman63,Weinberg65,ZeldovichGrishchuk}.
However, infinities generated by gravitational radiative processes
cannot be dealt with usual procedures, like renormalisation,
and call for the development of a new Quantum Gravity Theory
\cite{t'HooftVeltman}.
In the absence of a complete and consistent theory, discussions of
the effects of quantum fluctuations of gravitational field
have remained preliminary and very partial.
{}For the same reason, different approaches rely on specific
models or assumptions and they largely
differ in the way to introduce gravitational quantum fluctuations.

Most approaches consider that the Einstein equations
must still describe the metric structure of space-time at low energies
whereas profound modifications are expected for the ultimate
structure of space-time at very short distances,
due to quantum fluctuations of gravitation  \cite{Wheeler57}.
It has been speculated that these modifications, whatever their
precise form may be, should allow to remove the ultraviolet divergences
of Quantum Field Theory \cite{deWitt64,Isham71}.
Some remarkable features, which are already present but remain exceptional in
classical level, have been argued
to become important at energies comparable to Planck energy. The
representation of quantum evolution by means of path integrals has favored
developments analysing the effects of particular configurations of
classical metric, like black holes or more general topological
defects, on the nature of space-time at very short distances \cite{Hawking87}.
Approximate descriptions of a fluctuating space-time
have also been proposed, aiming to provide mechanisms producing
corrections to standard Quantum Mechanics,
localisation of classical objects or macroscopic decoherence
\cite{Karolyhazy66,EllisMohantyNanopoulos,Percival95}.

Attention will be focussed in the following on a different approach
which relies on a few minimal physical asumptions, deeply rooted
in the conception of a
space-time related to the properties of quantum vacuum.
Precisely, the Einstein equation will be considered as representing
the effective low energy behaviour of an underlying quantum theory.
The underlying theory may have quite different forms in the limit
of high frequencies where the quantum effects of gravity are expected
to become predominant. These differences will not be relevant
in the forthcoming discussions which depend only upon the general
relations between response functions and vacuum fluctuations.
Since these relations are used only at low frequencies lying much below
the Planck scale, significant results can be obtained
despite of the still unsolved problems of Quantum Gravity and, especially,
of the problems of renormalisability \cite{t'HooftVeltman}
and of vacuum stability \cite{HorowitzWald,HartleHorowitz}.
Note that attempts have been made to explain gravitation itself
as the result of perturbations of vacuum energy
\cite{Sakharov67,Adler82}.
In any case, quantum corrections to gravitational equations
are expected to take place and they may be studied with similar assumptions
\cite{Donoghue94,Ford95}.
The approach just outlined allows to derive universal noise spectra
for gravitational quantum fluctuations \cite{JaekelReynaud94}.
It thus leads to a quantum structure for space-time which does
not rely on any specific assumption and, in particular, does not
depend on any model dependent parameter.

It has been shown from minimal physical asumptions
that a quantum field theory describing gravitation by the exchange of
spin 2 particles coupled to the energy-momentum tensor leads
to the Einstein equation \cite{Feynman63,Weinberg65}. This equation thus
constitutes a firm basis for describing the propagation of gravitational
fields, and hence the properties of the associated vacuum fluctuations.
Because of the universal coupling of all fields to gravitation,
the gravitational vacuum also affects the vacua of all other fields.
The true quantum version of empty space is in fact the vacuum of quantum fields
coupled to gravitational vacuum. It is therefore important that gravitational
fluctuations and propagators are connected in the same manner
as for any other field theory.

Proceeding in this spirit, gravitational fluctuations are described as
perturbations $h_{\mu \nu }$ of Minkowski metric $\eta _{\mu \nu }$.
They are given an intrinsic description, independent
of a particular choice of reference system, when written in terms of
space-time curvatures. The Riemann tensor $R_{\lambda \mu \rho \nu }$
is invariant under gauge transformations, i.e. changes of coordinates.
It is defined in the Fourier domain at first order in the metric
perturbation as
\begin{equation}
R_{\lambda \mu \rho \nu }[k]={\frac 12}(k_\mu k_\nu h_{\rho \lambda
}[k]+k_\rho k_\lambda h_{\mu \nu }[k]-k_\mu k_\rho h_{\nu \lambda }[k]-k_\nu
k_\lambda h_{\mu \rho }[k])
\end{equation}
The quantum fluctuations of gravitational field are then obtained
from the free propagator associated with the linearised Einstein equation.
When written in terms of curvatures,
the corresponding noise spectrum takes a gauge independent form
\begin{eqnarray}
C_{R_{\lambda \mu \rho \nu }R_{\lambda ^{\prime }\mu ^{\prime }
\rho ^{\prime}\nu ^{\prime }}}[k]&=&
16\pi ^2 l_P^2 \theta (k_0) \delta (k^2)
\nonumber \\
&\times& ({\cal R}_{\lambda \mu \lambda ^{\prime }\mu ^{\prime }}
{\cal R}_{\rho\nu \rho ^{\prime }\nu ^{\prime }}
+{\cal R}_{\lambda \mu \rho ^{\prime }\nu^{\prime }}
{\cal R}_{\rho \nu \lambda ^{\prime }\mu ^{\prime }}
-{\cal R}_{\lambda \mu \rho \nu }
{\cal R}_{\lambda ^{\prime }\mu ^{\prime }\rho^{\prime }\nu ^{\prime }})
\nonumber \\
{\cal R}_{\lambda \mu \rho \nu }&=&{\frac 12}
(k_\lambda k_\rho \eta _{\mu \nu}+k_\mu k_\nu \eta _{\lambda \rho }
-k_\mu k_\rho \eta _{\lambda \nu}-k_\lambda k_\nu \eta _{\mu \rho })
\end{eqnarray}
These expressions correspond to gravitational wave zero-point fluctuations
and they can be seen to be determined by Lorentz invariance of vacuum,
symmetries of Riemann tensor and the vanishing of Einstein's tensor.
Their status is close to that of stochastic gravitational waves
which are predicted to have been generated by various cosmological
or astrophysical processes  \cite{Grishchuk77}.
As for vacuum fluctuations of the electromagnetic field,
these are irreducible fluctuations,
the spectral energy density of which
is normalised to $\hbar \omega /2$ per mode  \cite{Grishchuk90}.

The curvature tensor determines the gravitational effect on field
propagation. The momentum of any field
follows a law of geodesic deviation, and thus integrates curvature
perturbations encountered along propagation. This geodesic deviation
takes the same expression as the frequency shift induced by a classical
background gravitational wave
\cite{SachsWolfe,BraginskyMensky,MashhoonGrishchuk}.
When integrated along the propagation of the field, curvature fluctuations
induce fluctuations of propagation distances.
These geodesic fluctuations affect space-time localisation, which uses probe
fields,
as well as the definition of vacuum of quantum fields.
{}For space-time localisation, the fluctuations induced on the probe field
result in a noise spectrum for measured lengths or positions
\cite{JaekelReynaud94}.
For frequencies higher than the inverse of the propagation time, the noise
spectrum
is shown to take a universal form
\begin{equation}
C_{qq}[\omega ]\simeq l_P^2{\frac{\theta (\omega )}\omega }  \label{vdf}
\end{equation}
with a further numerical factor depending on the particular measurement
technique used, for instance one-way or round trip probing.
Curvature fluctuations then impose a limit on space-time probing of the
order of Planck length. The spectrum of fluctuations in vacuum (\ref{vdf}),
which holds for frequencies smaller than Planck frequency $\frac{c}{l_P}$ has
been
deduced from the minimal assumptions discussed previously, namely
an effective behavior of gravitation at low frequencies given by
Einstein equation, and a conformity of vacuum fluctuations with
fluctuation-dissipation relations. This spectrum exhibits the characteristic
properties of vacuum correlations, with fluctuations (\ref{vdf}) restricted
to positive frequencies. Hence, a commutator spectrum is obtained from
(\ref{vdf}),
which implies that geodesic distances are non-commutative
and behave as quantum variables.
Thus, the minimal assumptions just mentioned lead to behaviours
characterising an underlying non-commutative geometry.
This pleads for a drastic renouncement to the classical nature of space-time
and, accordingly, to the associated classical geometry.
Note that quantum generalisations of geometry have already been proposed
where the symmetry groups associated with classical
geometry are replaced by quantum groups
\cite{Woronowicz87,Majid88,LukierskiRuegg,Connes95,Kempf97}.
It has to be emphasized that the predictions of the approach sketched
here does not depend on any free parameter.
They may hence be used to select among the many candidates those
generalisations of classical geometry which fit the
Einstein equation while making gravitational vacuum fluctuations
conform to fluctuation-dissipation relations.

{}Fluctuations of geodesic distance induced by vacuum gravitational waves
take the same form as fluctuations of position (\ref{cqqev})
induced by radiation pressure of electromagnetic vacuum. The
order of magnitude is however different since it is here determined by the
Planck
length and not by the Compton wave-length. In particular, fluctuations of
geodesic
distances do not depend on the mass of the mirror.
In a position measurement performed with a probe field,
two different regimes of ultimate quantum noise are found,
depending on the endpoint mass used. For a ``microscopic'' mass,
smaller than the Planck mass, radiation pressure fluctuations dominate.
But for a ``macroscopic'' mass, greater than the Planck mass,
the ultimate quantum noise no longer depends on the test mass used and
must be attributed to fluctuations of space-time itself.
Geometric fluctuations of quantum space-time are probed only
when macroscopic endpoint reflectors are used. These results
suggest that a natural borderline might be delineated between
the macroscopic and microscopic world provided that quantum
fluctuations of gravity are accounted for.

We have discussed in the present section
how massless fields allow to probe light-like space-time intervals.
Massive probe fields can also be used, as is the case for instance in atomic
interferometers \cite{Borde97}. They would allow
to test time-like intervals \cite{JaekelReynaud95c}.
We may also note that effects of gravitation
vacuum fluctuations subsist for quantum fields
which are themselves in vacuum. Indeed, vacuum field correlations depend
on the geodesic distance which characterize the relative positions
of two points in space-time.
On the other hand, vacuum field fluctuations carry energy-momentum
fluctuations which are themselves a cause for metric fluctuations.
As a result, the gravitational and non-gravitational vacua must be
treated as coupled systems \cite{JaekelReynaud95c}.

\section{Conclusion and perspectives}

We have seen that vacuum fluctuations have a profound impact
on fundamental concepts of mechanics and of relativity.
In particular, the principle of relativity of motion is directly
connected to the symmetries of vacuum.
This connection was well-known for the cases of uniform velocity
and Lorentz invariance and it has to be extended to uniform acceleration
and conformal symmetry.
As soon as non uniform acceleration is involved, motion in vacuum gives
rise to radiation reaction forces opposed to the motion.

In order to appreciate the significance of these results,
it is worth recalling the difference between the covariance properties
which constitute the basement of General Relativity
and the invariance properties which are at the heart of the Special Theory
of Relativity \cite{Norton93}.
Clearly, the connection between vacuum symmetries and relativity of motion
lies on the side of invariance properties.
To go further in the same direction,
it may be considered that this connection challenges the general principle
of relativity, that is the principle of relativity of arbitrary motion.
At this point, two different opinions may be upheld.
On one hand, it may be hoped that a future theory
of Quantum Gravity will cure this difficulty and revive the general
principle of relativity at the quantum level.
On the other hand, it may be considered that uniform velocity and
uniform acceleration effectively play a privileged role since they
represent the only frictionless motions in vacuum.
In any case, the concept of relativity of motion should play a key role in
the development of theoretical frameworks consistently dealing
with quantum fluctuations and gravitational phenomena.

The same question also plays a central role in the construction
of a concept of space-time accounting for quantum fluctuations
as well as for relativistic symmetries.
As well-known since the advent of Relativity Theory, time
and space can no longer be considered as absolute notions
as it was the case in the Newtonian framework.
The basic element of a relativistic conception of space-time is that
of events occuring at some position in space and time \cite{Einstein}.
Comparison of events occuring at different
locations is performed through clock synchronisation
procedures built on the transfer of electromagnetic signals.
These ideas have now been included in metrological definitions
of time and space \cite{TimeFrequency}.
At some high level of precision, synchronisation procedures
have to reach a limit associated with the quantum nature
of the signals used in the transfer \cite{NewtonWigner,SaleckerWigner}.
More profoundly, space-time observables associated with a given event
certainly belong to the quantum domain, like atomic clocks used for time
definition and electromagnetic signals used for synchronisation.
As a consequence, the
shifts of space-time observables under transformations to accelerated frames
cannot be deduced from covariance properties associated
with the corresponding map transformations.
It has recently been shown that these questions may be given
satisfactory answers by using the same
symmetries which have been used here to represent the properties of motion.
The conformal algebra indeed determines the conserved quantities.
{}For field states with a non vanishing mass, such as those used to perform
the Einstein localisation of an event in space-time, these quantities may be
used
to define quantum observables associated with positions of an event
in space-time \cite{JaekelReynaud96}.
The conformal algebra also fixes the shift
of these observables under changes of reference frames
and thus describes within a quantum framework the
relativistic effects associated with uniform and accelerated motions.
The shift of mass under transformations
to accelerated frames may be written in terms of position observables,
in consistency with Einstein principle of equivalence
\cite{JaekelReynaud97}.

Space-time observables conceptually
differ from coordinate parameters, as emphasized by their
relativistic and quantum properties.
In particular, a change of reference frame has to be distinguished from a
change of coordinate map, since the former is directly related to relativistic
symmetries whereas the second is a mere matter of convention.
The problem is particularly acute for the definition of time
which has been thoroughly discussed in connection
with fundamental questions of Quantum Measurement Theory
\cite{Jammer74} and Quantum Gravity \cite{Isham93}.
A time observable associated with the position of an event
is defined, besides space observables, by the procedure just described.
It is however a localisation observable which differs from any kind of
evolution parameter.
A lasting challenge is hence to represent movement in a quantum framework
where the prime roles are played by conserved quantities
\cite{PageWootters,Unruh89,Rovelli9195,JaekelReynaud97b}.
It is amazing that this paradox of modern physics lies in the continuity
of the logical debates of the ancient philosophers.
It might be that its solution requires a new definition of movement which
should
be conceived, in the spirit of the atomistic line of thought, as a
sequence of events \cite{RussellRelat}.

\vspace{5mm}

\noindent{\bf Acknowledgements}
Thanks are due to Vladimir Braginsky, Jean-Michel Courty,
Leonid Grishchuk, Astrid Lambrecht, Paolo
Americo Ma\"{\i}a Neto, Roberto Onofrio, Carlo Rizzo and Lorenza Viola
for stimulating discussions.

\end{document}